
\documentstyle{amsppt}
\magnification 1200
\NoBlackBoxes
\NoRunningHeads

\topmatter
\title Algebraic integrability of Schr\"odinger operators
       and representations of Lie algebras
  \endtitle
\author {\rm {\bf Pavel Etingof and Konstantin Styrkas} \linebreak
	\vskip .1in
   Department of Mathematics\linebreak
   Yale University\linebreak
   New Haven, CT 06520, USA\linebreak
   e-mail: etingof\@math.yale.edu,
			styrkas\@math.yale.edu}
\endauthor

\endtopmatter

\define\dd{\partial}
\define\a{\alpha}
\define\la{\lambda}
\define\ver{M_\lambda}
\define\dual{M_\lambda^*}

\define\DP{\Delta^+}

\define\tensor{\hat \otimes}

\define\twin{\Phi_\lambda^u}
\define\tw{\tilde\Phi_\lambda^u}
\define\td{t\cdot f_}

\define\hw{v_\lambda}
\define\lw{v_{-\lambda}^*}

\define\mlt{{(\bold n)}}
\define\End{End(U[0])}

\define\({\left(}
\define\){\right)}
\define\<{\langle}
\define\>{\rangle}
\define\g{\frak g}
\define\h{\frak h}
\define\np{\frak n_+}
\define\nm{\frak n_-}
\define\Z{{\Bbb Z}}
\define\N{{\Bbb N}}

\define\C{{\Bbb C}}
\define\lp{\la^\perp}
\define\hot{{\{\it higher order terms\}} }
\define\ldt{{\{\it lower degree terms\}} }
\define\KK{\<\a,\la+\rho\>-\frac n2\<\a,\a\>}

\document
\centerline{March 18, 1994}
\centerline{hep-th 9403135}
\vskip .1in
\centerline{\bf 1. Introduction}
\vskip .1in

 A Schr\"odinger operator is a differential operator whose symbol
is the Laplace's operator. A quantum integral of a Schr\"odinger
operator is a differential operator that commutes with it.

A Schr\"odinger operator in $m$ variables is called integrable
if it has $m$ algebraically independent quantum integrals
in involution (i.e. commuting with each other).
This notion is the quantum analogue of the notion of Liouville
integrability of a classical Hamiltonian system.

A Schr\"odinger operator in $m$ variables is called algebraically
integrable if it is integrable but the algebra of its quantum integrals cannot
be generated by $m$ operators. In the one-variable case,
algebraically integrable operators correspond to finite-gap potentials
\cite{Kr}.

One of the most interesting examples of an integrable
Schr\"odinger operator is the Calogero-Sutherland operator
\cite{C},\cite{S},\cite{OP}.
This is the Hamiltonian of the quantum many-body
problem with rational, trigonometric, or elliptic
interaction potential. The Calogero-Sutherland operator
depends on a parameter which is called the coupling constant.

It has been observed \cite{CV1},\cite{CV2},\cite{VSC}
 that the Calogero-Sutherland
operators become algebraically integrable when
the coupling constant takes a discrete set of special values.
This is proved for the rational and trigonometric case
but still remains a conjecture in the elliptic case
for two or more variables.

These results can be generalized to Calogero-Sutherland operators
associated with root systems, which were defined in
\cite{OP}.

In this paper we study integrability and algebraic integrability
properties of certain matrix Schr\"odinger operators.
More specifically, we associate such an operator (with rational,
trigonometric, or elliptic coefficients) to every
simple Lie algebra $\g$ and every representation $U$ of this algebra
with a nonzero but finite dimensional zero weight subspace.
(The Calogero-Sutherland operator is a special case of this
construction). Such an operator is always integrable \cite{E}.
Our main result is that it is also algebraically integrable
in the rational and trigonometric case
if the representation $U$ is highest weight. This generalizes
the corresponding result for Calogero-Sutherland operators
(\cite{CV1}). We also conjecture that this is true for the elliptic
case as well, which is a generalization of the corresponding
conjecture from \cite{CV2}.

The proof of the main result is based on the method of
$\psi$-function -- a joint eigenfunction
of quantum integrals of the Schr\"odinger operator.
This method was developed in \cite{CV1}.
The proof of existense and uniqueness of the $\psi$-function
is based on an explicit construction of this function
which uses representation theory of the Lie algebra $\g$.
To be more precise, the $\psi$-function is realized
(up to a factor) as a weighted trace of an intertwining
operator between a Verma module over $\g$ and the tensor product
of this module with $U$. Such realization goes back to
\cite{E},\cite{EK1}, where it is found that joint eigenfunctions
of quantum integrals of a Calogero-Sutherland operator
can be realized as traces. Using the theory
of Shapovalov form for $\g$ (\cite{Sh},\cite{KK}),
we prove that the trace function
satisfies the axioms for the $\psi$-function analogous to those
formulated in \cite{CV1}, and is determined uniquely by them,
and then establish algebraic integrability using the method of
\cite{CV1}.

The paper is organized as follows. In Section 2 we make
the necessary definitions, motivate them,  and formulate the main result.
In Section 3 we give information about Verma modules, Shapovalov
form, and intertwining operators. In Section 4 we define the
$\psi$-function as a normalized trace, and prove two properties
of this function. In Section 5 we prove that these two properties
uniquely determine the $\psi$-function. In Section 6 we prove
algebraic integrability using the $\psi$-function. In the Appendix
we describe how to get Weyl group invariant quantum integrals
from central elements (Casimirs) of $U(\g)$.
\vskip .1in

{\bf Acknowledgements. } The authors would like to thank
I.Frenkel, A.Kirillov Jr., F.Malikov and A.Veselov
for numerous discussions and useful suggestions.

The first author's work was supported by the Alfred P.Sloan
Graduate Dissertation Fellowship.
\vskip .1in

%

\centerline{\bf 2. Main definitions and results}
\vskip .1in

Let $V$ be a finite-dimensional complex vector space.

\proclaim{Definition 2.1}
A matrix differential operator is a differential
operator whose coefficients are $\text{End}V$-valued functions.
A matrix Schr\"odinger operator is a differential operator
of the form
$$
L=\Delta-A(x),\ x\in\C^m,\tag2-1
$$
where $\Delta$ is the Laplacian in $\C^m$, and $A$ is a
meromorphic function in $\C^m$ with values in
$\text{End}V$.
\endproclaim

\proclaim{Definition 2.2} A matrix Schr\"odinger operator $L$
is called integrable if there exist pairwise commutative
matrix differential operators $L_1=L,L_2,...,L_m$ such that
the symbols of $L_i$ have the form $p_i(\frac{\dd}{\dd
x_1},...,\frac{\dd}{\dd x_m})\text{Id}$, where $p_i$ are algebraically
independent polynomials ($p_1(\bold y)=\bold y^2$).
The operators $L_1,...,L_m$ are called the quantum integrals
for $L$.
\endproclaim

Let $R\subset \frak h^*=\C^m$
be the root system of a simple Lie algebra $\g$
of rank $m$, and let $\Delta^+$ be the set of
positive roots of $R$.

\proclaim{Definition 2.3} The Calogero-Sutherland (CS) operator
for $R$ is the operator
$$
L=\Delta-\sum_{\alpha\in\Delta^+}C_{\alpha}u(<x,\alpha>)+K,\tag2-2
$$
where the scalar constant $C_{\alpha}$ may depend only on the length of
the root $\alpha$, and $u$ is one of the following potential
functions: (i) $u(x)=2/x^2$ (rational potential), (ii) $u(x)=2/sinh^2x$
(trigonometric potential),
or (iii) $u(x)=2\wp(x|\omega_1,\omega_2)$ (elliptic potential), where
$\wp(x|\omega_1,\omega_2)$ is the Weierstrass elliptic function with periods
$\omega_1,\omega_2$, and $K$ is a constant
 (Cases (i) and (ii) are degenerations of case (iii)).
\endproclaim

Such operators were introduced by Calogero \cite{C} and Sutherland
\cite{S} for the root system $A_m$ and by Olshanetsky and Perelomov
\cite{OP} in general.

\proclaim{Theorem 2.1}  The operator $L$ given by (2-2)
is integrable. Furthermore, the symbols
of the quantum integrals $L_i$, $1\le i\le m$, generate the algebra
of Weyl group invariant polynomials on $\frak h$.
\endproclaim

For $R=A_m$ (and in some other special cases) this theorem
was proved in \cite{OP}. Cases (i) and (ii) for general root systems
were settled by Heckman and Opdam \cite{HO,H1,O1,O2}.
Case (iii) for $B_m$, $C_m$ and $D_m$ is
settled in \cite{Osh}. The general proof for Case (iii)
(and hence Cases (i) and (ii)) was given recently by
I.Cherednik \cite{Ch}.

If $m=1$ then any Schr\"odinger operator is integrable by the definition.
In two or more variables integrability is a very rare
property. This is illustrated by the following result.

\proclaim{Theorem 2.2} \cite{OOS,OS} Let $m\ge 2$. Let $L$ be an integrable
Schr\"odinger operator defined by (2-1) with $V=\C$. Assume that
the operators $L_i$, $1\le i\le m$, are invariant under the symmetric group
$S_{m+1}$ acting irreducibly in $\C^m$, and their symbols generate
the ring of $S_{m+1}$-symmetric polynomials on $\C^m$ (the operator
$L_i$ is of order $i+1$).
Then $L$ coincides with (2-2) for the root system $A_m$
for some values of the parameters.
\endproclaim

Theorem 2.1 can be generalized to the matrix case, as follows.

Let $\g$ be a complex simple Lie algebra, $\frak h\subset\g$
be a Cartan subalgebra, $R\subset\frak h^*$ be the root system of
$\g$, $\Delta^+$ be the set of positive roots.
$e_{\alpha},f_{\alpha}\in\g$ be the root elements corresponding
to the positive root $\alpha$. Let $U$ be a representation
of $\g$ such that the space $V=U[0]$ of zero weight vectors in $U$
is finite-dimensional. Define the matrix Schr\"odinger operator
$$
H_{\g,U,u}=\Delta_{\frak h} -\sum_{\alpha\in\Delta^+}
u(<x,\alpha>)e_{\alpha}f_{\alpha}, \ x\in\frak h,\tag
2-3
$$
where $u$ is of type (i), (ii), or (iii) from Definition 2.3.
Such operators are considered in \cite{E},\cite{EK1}.

\proclaim{Theorem 2.3} The operator (2-3) is integrable.
The symbols of its quantum integrals are generators
of the algebra of Weyl group invariant polynomials on $\frak h$.
\endproclaim

This result is proved in \cite{E} for the special case
$\g=\frak{sl}_{m+1}$, but the method used for the proof
works for any Lie algebra. This method uses representation
theory of the Lie algebra $\g$ in the trigonometric case,
and representation theory of the affine Lie algebra $\hat\g$
in the elliptic case. The quantum integrals of $H_{\g,U,u}$
are constructed from central elements of the universal
enveloping algebra. We discuss this method in the Appendix.

As a particular case, Theorem 2.3 includes Theorem 2.1
for the root system $A_m$. Indeed, let us take $\g=\frak{sl}_m$ and a
special representation of $\g$:
$U_{\mu}=(z_1...z_{m+1})^{\mu}\C[\frac{z_1}
{z_2},...\frac{z_m}{z_{m+1}},\frac{z_{m+1}}{z_1}]$, $\mu\in\C$, with
the action of $\g$ by linear transformations of variables
(this representation has no highest weight).
All weight subspaces in $U_\mu$ are one-dimensional; in particular,
$V=U_\mu[0]=\C$. It is easy to compute that
$e_{\alpha}f_{\alpha}|_{U[0]}=\mu(\mu+1)$. Therefore,
if $\mu$ is chosen in such a way that $C_{\alpha}=\mu(\mu+1)$,
then operator (2-3) transforms into (2-2) for the root system $A_m$.

Krichever \cite{Kr} introduced the notion of an algebraically
integrable Schr\"odinger operator (see also \cite{CV1}).
Here we generalize this definition
to the matrix case.

Let $L$ be an integrable matrix
Schr\"odinger operator, let $L_1=L,...,L_m$ be its quantum integrals,
and let the symbols of $L_i$ be $p_i(\dd/\dd x_1,...,\dd/\dd x_m)
\text{Id}$, where $p_i$ are algebraically independent polynomials.

\proclaim{Definition 2.4}  $L$ is
called algebraically integrable if there exists a matrix differential
operator $L_0$ commuting with $L_1,..., L_m$
with symbol $p_0(\dd/\dd x_1,...,\dd/\dd x_m)\text{Id}$, such that
for generic $E_1,...,E_m\in\C$ the polynomial
$p_0$ takes distinct values at the roots of the system of equations
$p_i(y_1,...,y_m)=E_i$, $1\le i\le m$.
\endproclaim

For $V=\C$ this definition coincides with the one in \cite{Kr},\cite{CV1}.
In the matrix case and $m=1$ the property of algebraic integrability
of differential operators was studied in \cite{G}.

It turns out that a Calogero-Sutherland operator is algebraically
integrable for a discrete spectrum of values of the constants
$C_{\alpha}$.

\proclaim{Theorem 2.4}\cite{VSC} If $C_{\alpha}=
\frac{1}{2}\mu_{\alpha}(\mu_{\alpha}+1)<\alpha,\alpha>$ for all roots
$\alpha\in\Delta^+$, where $\mu_{\alpha}$ is an integer depending
only on the length of $\alpha$, then the operator (2-2)
is algebraically integrable for the rational and
trigonometric potential.
\endproclaim

\proclaim{Conjecture 2.5}\cite{CV2} Theorem 2.4 is true
for the elliptic potential.
\endproclaim

Conjecture 2.5 is proved only for the case of the root system $A_1$.
In this case, operator (2-2) is the Lam\'e operator
$L=\dd^2-C\wp$, and algebraic
integrability of this operator is equivalent
to the finite gap property, which takes place
for $C=\mu(\mu+1)$, $\mu\in\Z$ \cite{Kr}.
In this case, there is a quantum integral $L_0$
of order $2\mu+1$.

Now let us consider the case of the root system $A_m$.
Looking at the interpretation of the Calogero-Sutherland operator
via the representation $U_{\mu}$, we see why the integer values
of $\mu$ should be special: they are exactly those values
for which the representation $U_{\mu}$ has
a finite dimensional submodule or quotient module which is
isomorphic to a symmetric power of $\C^{m+1}$ (or $(C^{m+1})^*$).
Since the zero weight vector is contained in
this finite-dimensional module,
we can use it instead of $U_{\mu}$.
Thus we observe that algebraic integrability occurs at those values of $\mu$
where $U_{\mu}$ can be replaced by a highest weight module.
This motivates the following general theorem which is the main result
of this paper.

\proclaim{Theorem 2.6} If $U$ is a highest weight $\g$-module then
$H_{\g,U,u}$ is algebraically integrable for the rational and
trigonometric potential.
\endproclaim

In Sections 3-6
we prove this theorem for the trigonometric case. The rational case
can be obtained in the limit, so we don't discuss it.

Note that Theorem 2.4 for the root system $A_m$
is a special case of Theorem 2.6.

Finally, we would like to formulate a natural conjecture concerning the
elliptic case.

\proclaim{Conjecture 2.7} Theorem 2.6 is true
for the elliptic potential.
\endproclaim

This conjecture contains Conjecture 2.5 for the root system $A_m$.
We believe that it could be proved by applying the methods
of this paper to the elliptic case and using the techniques of
 representation theory of affine Lie algebras and theory
of vertex operators introduced in \cite{E},\cite{EK1}.
\vskip .1in


\vskip .1in
\centerline{\bf 3. Verma modules, Shapovalov form, intertwining operators.}
\vskip .1in

Let $\g$ be a simple complex Lie algebra with triangular decomposition
$\g=\nm\oplus\h\oplus\np$. Fix an element $\la \in \h^*$.
Denote by $\ver$ the Verma module over $\g$ with highest weight $\la$,
i.e. the module with one generator $\hw$ and relations
$$\np\hw=0, \qquad h\hw = \<\la,h\>\hw \qquad \text{for } h\in\h.$$

We have the decomposition $\ver=\bigoplus_{\mu\le\la} M_\la[\mu]$ of $\ver$
into the direct sum of finite dimensional weight subspaces $M_\la[\mu]$.
Denote also $\dual=\bigoplus_{\mu\le\la}M_\la[\mu]^*$ the restricted
dual module
to $\ver$ with the action of $\g$ defined by duality. $\dual$ is a
lowest weight module with the lowest weight vector $\lw$ of weight $-\la$.

We have a vector space decomposition $U(\g)=U(\nm)\otimes U(\h)\otimes
U(\np).$ Define the Harish-Chandra homomorphism
$\phi:U(\g)[0] \to U(\h)$ by
$\phi|_{U(\h)}=Id,$  and $\phi(g)=0$ if $g\in U(\g)[0]$ can be
represented as $g=g_1e_i$ for some $g_1\in U(\g), e_i\in \np.$
This in turn gives rise to a contravariant bilinear $U(\h)$-valued
form $\Cal F$ on $U(\nm)$ defined by
$$\Cal F(g_1,g_2)=\phi(\omega(g_1) g_2),$$
when $g_1, g_2$ belong to the same weight subspace of $U(\nm)$,
and $\Cal F(g_1,g_2)=0$ otherwise. Here $\omega$ is the Cartan
antiautomorphism of $\g$ defined by
$$\omega(e_i)=f_i,\quad\omega(f_i)=e_i,\quad\omega(h_i)=h_i$$
It is easy to see that this form is symmetric.

As $U(\h)$ can be identified with the space of all polynomials on
$\h^*,$ we can introduce a symmetric contravariant $\C$-valued form
$F$ on $\ver$ defined by
$$F(g_1\hw,g_2\hw)=\Cal F(g_1,g_2)(\la).$$

Let $U$ be any weight $\g$-module with finite dimensional
zero weight space $U[0]$. The completed tensor product
$\ver\hat\otimes U = Hom_\C(\dual,U)$ has a natural $\g$-module
structure. We say that an element $v\otimes u$ has order $\eta$ if
$v\in M_\la[\la-\eta]$. Clearly, only elements of order $\eta\in Q^+$
may occur, where $Q^+=\sum_{\a\in\DP}\Z_+e_\a.$ We say that
$\nu<\eta$ if $\nu\ne\eta$ and $\eta-\nu\in Q^+.$

Let $u\in U.$ Let $\twin:\ver\to\ver\tensor U$
be an intertwining operator such that \linebreak
$\hw\mapsto\hw\otimes u + $\hot.

It is clear from the intertwining property of $\twin$ that $u$ has
to be a zero weight vector.

\proclaim{Proposition 3.1}
\roster
\item If $\ver$ is irreducible then $\twin$ exists and is unique
for any $u\in U[0]$.
\item If $\ver$ is reducible then $\twin$ exists iff $u\in U[0]$ is such
that $I_\la u = 0,$ where $I_\la\subset U(\np)$ is the left ideal
consisting of all $g\in U(\np)$ such that $g\lw = 0.$
\endroster
\endproclaim

\demo{Proof}
First consider the case when $\ver$ is irreducible. Because $\ver$ is
freely generated by $\hw$ over $U(\nm)$, we only need to prove that the
module $\ver\tensor U$ contains a unique singular vector of the form
$\hw\otimes u +$\hot. This is the same as to construct a map
$\Theta:\dual\to U$ such that $\Theta(\lw)=u$ and
$\Theta$ is a $\np$-intertwiner. But $\dual$ is a free $U(\np)$-module
generated by $\lw$, so $\Theta$ can be uniquely extended from $\lw$ to
the whole $\dual$.
Now let $\ver$ be reducible. Then $\dual$ is also reducible, and the
universal enveloping algebra $U(\np)$ contains the nonzero annihilating ideal
$I_\la$ of the vector $\lw$. The same argument as above shows that
$I_\la u=0$ is a necessary and sufficient condition for the existense
of the map $\Theta$. If it is satisfied we can put $\twin(\hw)$
to be the singular vector of weight $\la$ in $\ver\tensor U$, and then
extend this map to the whole $\ver$. \qed
\enddemo

It is known that $\ver$ is irreducible for generic $\la$. For special
$\la$'s $\ver$ may be reducible, and it happens when the
contravariant bilinear form on $\ver$ is degenerate. Shapovalov
\cite{Sh} obtained
an explicit formula for the determinant of this form:
$$\det F_\mu(\la)=
const\prod_{\a\in\DP}\prod_{n\in\N}(\KK)^{K(\mu-n\a)},\tag3-1$$
where $F_\mu = {(F_\mu)}_{i,j}, \quad i,j = 1,2,\dots, \dim M_\la[\la-\mu]$ \
is the matrix of the restriction of the form to $M_\la[\la-\mu]$, $K(\mu)$ --
the Kostant partition function, and the nonzero constant depends on
the choice of basis in $M_\la[\mu].$

Let $$\chi_n^\a(\la)=\KK.\tag3-2$$
The conditions for reducibility of $\ver$ can
then be rewritten as
$$\chi_n^\a(\la)=0 \text{ for some } \a\in\DP, n\in\N$$
Now we fix weight $\mu$ and let $n_\mu^\a=\max\{n\in\N|K(\mu-n\a)\ne0\}$.
Denote
$$\chi_\mu(\la) = \prod_{\a\in\DP}\prod_{n=1}^{n_\mu^\a}\chi_n^\a(\la)\tag3-3$$

We need the following
\proclaim{Lemma 3.2}
Matrix elements ${(F_\mu^{-1})}_{i,j}$ of the inverse matrix
$F_\mu^{-1}$ can be written in the form
$${(F_\mu^{-1})}_{i,j} = \frac {P_{ij}^\mu(\la)}{\chi_\mu(\la)}$$
for some suitable polynomials $P_{ij}^\mu(\la)$.
\endproclaim

\demo{Proof}
Shapovalov formula implies that matrix elements are rational functions
in $\la$ with only possible poles in hyperplanes defined by $\chi_n^\a(\la)=0.$
Our goal is to show that only simple poles may occur.

Fix $\a\in\DP,n\le n_\a.$ Take $\la$ such that $\chi_m^\beta(\la)=0$
iff $m=n, \beta=\a.$ Then $\ver$ is reducible and contains a unique
maximal submodule of $\ver^1,$ generated by a singular vector $v_{\la-n\a}.$

Fix $z\in\h^*$ such that $\<\a,z\>\ne0$ for any $\a\in\DP,$ and let
$t$ be an independent variable. Using the $U(\h)$-valued bilinear form
$\Cal F$ we can introduce a new $\C[t]$-valued bilinear form $F^t$ on
$\ver$ defined by
$$F^t(g_1\hw,g_2\hw)=\Cal F(g_1,g_2)(\la+tz),\quad g_1,g_2\in U(\nm)$$
Clearly, specialization $t\mapsto0$ gives the usual Shapovalov form.

Denote $N=\dim M_\la[\la-\mu]=K(\mu), M=\dim M_\la^1[\la-\mu]=K(\mu-n\a)$.
Choose a basis $v_k$ in $M_\la[\la-\mu]$ so that $\{v_i\}, i=1,2,\dots,M,$
would form a basis for $M_\la^1[\la-\mu].$ Then the matrix elements
${(F_\mu^t)}_{i,j}$ will be divisible by $t$ if $i\le M$ or $j\le M:$

$$F_\mu^t=\pmatrix
\td{1,1} & \dots & \td{1,M} & \td{1,M+1} & \dots & \td{1,N} \\
\vdots  &       & \vdots  & \vdots    &       & \vdots   \\
\td{M,1} & \dots & \td{M,M} & \td{M,M+1} & \dots & \td{M,N} \\
\\
\td{M+1,1} & \dots & \td{M+1,M} & f_{M+1,M+1} & \dots & f_{M+1,N} \\
\vdots  &       & \vdots  & \vdots    &       & \vdots   \\
\td{N,1} & \dots & \td{N,M} & f_{N,M+1} & \dots & f_{N,N}
\endpmatrix$$
where $f_{i,j}$ are some polynomials in $\la,t$.
It is clear now that the determinant of any $(N-1)\times(N-1)$
submatrix of $F_\mu^t$ is divisible by $t^{M-1}.$ Shapovalov formula
implies that $\det F_\mu^t$ is divisible by exactly $M$th power of $t$,
which means that when we compute the matrix elements of
${(F_\mu^t)}^{-1}$, only simple poles will be allowed when
$t=0$, or, equivalently, ${(F_\mu^{-1})}_{i,j}$, will have at most
simple poles on the hyperplanes $\chi_n^\a(\la)=0.$
Repeating this argument for all $\mu,n,\a$ we prove the lemma.\qed
\enddemo

We can apply this result to get more information about the
intertwining operator $\twin.$ In the proof of Proposition
we defined $\twin\hw$ as a map $\Psi:\dual\to U.$ We would like to
obtain a more explicit formula for $\twin\hw$ as an element of
$\ver\tensor U.$

For any basis $v_k=g_k^\mu\hw, g_k^\mu\in U(\nm)$ of $M_\la[\la-\mu]$
we have the basis of $\dual[-\la+\mu]$ given by $v_k^*=(\omega g_k^\mu)\lw,$
where $\omega$ is the Cartan involution. It is clear that
$\<v_i^*,v_k\>=F(v_i,v_k).$0z

Introduce another basis $w_k$ which is dual to $v_k^*$ in the usual
sense, i.e. $\<v_i^*,w_j\>=\delta_{ij}.$
These two bases $v_k$ and $w_k$ are related via the $F_\mu$ matrix:
$$ v_k=\sum_i {(F_\mu)}_{ki}w_i,\qquad\text{or}$$
$$ w_k=\sum_i{(F_\mu^{-1})}_{ki}v_i$$

It is clear that in this notation
$$\twin\hw=\hw\otimes u+\dots+\sum_k w_k\otimes(\omega g_k^\mu)u+\dots=
\hw\otimes u+\dots+\sum_{k,l} {(F_\mu^{-1})}_{kl}\cdot
 g_k^\mu\hw\otimes(\omega g_l^\mu)u+\dots$$

\proclaim{Corollary 3.3}
Suppose we have an intertwiner $\twin:\ver\to\ver\tensor U,$ where
$U$ is a highest weight $\g$-module with the highest weight $\theta.$
\roster
\item There are no order $\mu$ terms in the expression for $\twin\hw.$
unless $\mu\le\theta.$
\item If $\mu\le\theta$ then the order $\mu$ part of $\twin\hw$ can
be written as
$$\sum_{k,l}{(F_\mu^{-1})}_{kl}\cdot g_k^\mu\hw\otimes(\omega
g_l^\mu)u,\tag3-4$$
where
$${(F_\mu^{-1})}_{kl}=\frac{P_{kl}^\mu(\la)}{\chi_\mu(\la)}$$
for some polynomials $P_{kl}^\mu(\la).$
\item The condition of existence of $\twin$ can be written as
$\chi_\theta(\la)\ne0$.
\item for any basis $g_k^\mu$ of $U(\nm)$ we can choose polynomials
$S_{kl}^\mu(\la)$ so that
$$\twin\hw=\sum_{\mu\in
Q^+}\sum_{k,l}\frac{S_{kl}^\mu(\la)}{\chi_\theta(\la)}
g_k^\mu\hw\otimes(\omega g_l^\mu)u.\tag3-5$$
\endroster
\endproclaim

For a rational function $R,$ represented as a ratio of two polynomials
$R=\frac PQ$ we set $\deg R = \deg P - \deg Q.$

Note that all coefficients ${(F_\mu^{-1})}_{kl}$ are of negative
degree in $\la.$ Later we will work with $\la$ in the hyperplanes
$\<\a,\la\>=const,$ so we introduce notation
$$\la_\a=\frac{\<\a,\la\>}{\<\a,\a\>}\a, \quad \lp=\la-\la_\a,\tag3-6$$
so that $\lp$ is a $(\dim\h-1)$-dimensional vector and
$\<\a,\lp\>=0.$

We will use the following

\proclaim{Proposition 3.4}
When restricted to the hyperplane $\<\a,\la\>=C,$
matrix elements ${(F_\mu^{-1})}_{kl}$ are rational functions in
$\lp$ of nonpositive degree, and only constants may occur as terms of
degree 0.
\endproclaim

\demo{Proof}

We choose a special basis in $U(\nm)[\mu]$. For any
sequence $\omega$ of positive roots
$\beta_1\ge\beta_2\ge\dots\ge\beta_r,$ where
$\ge$ denotes now the lexicographical order, such that $\sum \beta_i=\mu,$
set $X_\omega=f_{\beta_1}\dots f_{\beta_r}.$
Set $\deg X_\omega=Card(\{k|\beta_k\ne\a\}).$
We also write $f_\beta=g_{-\beta}$, $e_\beta=g_\beta$ for a positive
root $\beta.$

The set of $X_\omega$'s is a basis in $U(\nm)[\mu].$ We also have
$$\deg \Cal F(X_{\omega_1},X_{\omega_2})(\la)\le
\frac{\deg X_{\omega_1}+\deg X_{\omega_2}}2.\tag3-7$$

Indeed, we can only raise the degree by commuting
some $e_\beta$ and $f_\beta$ for $\beta\ne\a$, which results in
the term $\<\beta,\la\>+const$, which is linear in $\lp.$
Note also that commuting with $e_\a$ or $f_\a$ will not increase
the total number of terms $e_\beta$ and $f_\beta$ for all $\beta\ne\a.$
Therefore, the maximal degree cannot be greater than half the original
number of terms $e_\beta$ and $f_\beta$, $\beta\ne\a.$
This proves formula (3-7).

The determinant of the form in the hyperplane $\<\la,\a\>=const$ is equal to
$$\det F_\mu(\la)= const\prod_{\beta\ne\a}
\prod_{n\in\N}(\<\beta,\la+\rho\>-\frac n2\<\beta,\beta\>)^{K(\mu-n\beta)},$$
where the constant depends on C.
Then $$N=\sum_{\beta\ne\a}\sum_{n\in\N} K(\mu-n\beta)$$
is the degree of the determinant
as polynomial in $\lp.$
By the same argument as in \cite{Sh}, from (3-7) it follows that the
$\lp$-degree of any minor of the Shapovalov
matrix cannot exceed $N$. Moreover, commuting with
$e_\a$ or $f_\a$ does not change the set of $\beta\mod\a$, and
therefore any term of degree exactly $N$ has highest term proportional
to that of the determinant, which proves the Proposition. \qed
\enddemo


\vskip .1in
\centerline{\bf 4. Matrix Trace, $\psi$-function and its properties.}
\vskip .1in

Fix a highest weight $\g$-module $U$ with highest weight
$\theta$ and finite dimensional zero weight space $U[0].$
Consider a new operator
$$\tw=\chi_\theta(\la)\twin \tag4-1$$
{}From Corollary 3.3 it follows that
$$\tw\hw=\sum_\mu \sum_{k,l} S_{kl}^\mu(\la) g_k^\mu\hw\otimes g_l^\mu u$$
This expression allows us to define $\tw$ even for $\la$ such that
$\chi_\theta(\la)=0$ where $\twin$ itself was not defined.
It is clear that $\tw$ is an intertwining operator for all $\la,$
and it has the property
$$\tw\hw=\hw\otimes\chi_\theta(\la)u+\text{\hot}. \tag4-2$$

Let $\Psi(\la,x)$ be a $End(U[0])$-valued function on $\h^*\times\h$
defined by
$$\Psi(\la,x)u=\frac{Tr|_{\ver}(\tilde\Phi_\la^u
e^x)}{Tr|_{M_{-\rho}}(e^x)} \tag 4-3$$

\proclaim{Proposition 4.1}
The $\Psi$-function defined above has the following properties:
\roster
\item
$$\Psi(\la,x)=e^{\<\la+\rho,x\>}\tilde P(\la,x),\tag 4-4$$
where $\tilde P(\la,x)$ is a $End(U[0])$-valued polynomial in $\la$ with
the highest term
$$\prod_{\a\in\DP}\<\a,\la\>^{n_\a}\cdot Id,$$
and we put for brevity $n_\a=n_\theta^\a.$
\item
If $\KK=0$ for some $\a\in\DP,n=1,2,\dots,n_\a$ then
$$\Psi(\la,x)=\Psi(\la-n\a)\tilde B_n^\a(\la),\tag4-5$$
for some $\tilde B_n^\a(\la)\in End(U[0]),$ which is rational
in $\la$ of nonpositive  degree, and only constant operators may
appear in it as degree zero terms.
\endroster
\endproclaim
\demo{Proof}
The first part is clear from the formula
$$Tr|_{M_\la}(\tilde\Phi_\la^u e^{x})=\sum_\mu
e^{\<\mu,x\>}Tr|_{M_\la[\mu]}(\tilde\Phi_\la^u)=
e^{\<\la+\rho,x\>}\sum_{\beta\in
Q^+}e^{\<-\rho-\beta,x\>}Tr|_{M_\la[\la-\beta]}(\tilde\Phi_\la^u)$$
and the fact that all the $Tr|_{M_\la[\la-\beta]}(\tilde\Phi_\la^u)$
are some combinations of $S_{kl}^\mu$'s, and therefore polynomials in
$\la.$ Their highest terms are obviously all equal to
$\prod_{\a\in\DP}\<\a,\la\>^{n_\a},$ so the highest term of
$P(\la,x)$ is equal to
$$\frac{\sum_{\beta\in Q^+}e^{\<-\rho-\beta,x\>}K(\mu)}{Tr|_{M_{-\rho}}(e^x)}
\prod_{\a\in\DP}\<\a,\la\>^{n_\a}\cdot
Id=\prod_{\a\in\DP}\<\a,\la\>^{n_\a}\cdot Id,$$

We now prove the second property of the $\Psi$-function. Let $\KK=0$
for some $\a\in\DP,1\le n\le n_\a,$ but $\<\beta,\la+\rho\>-\frac
m2\<\beta,\beta\>\ne0$ unless $\beta=\a, m=n.$ Set $\mu=n\a.$

{}From Corollary 3.3 it follows that $\tw\hw$ has no order $\nu$ terms
unless $\nu\ge n\a.$ In particular, there are no order zero terms. On
the other hand, $\tw\hw$ has to be a singular vector.  Therefore we
must have
$$\tw\hw=v_{\la-n\a}\otimes\tilde u+\text{\hot},\tag 4-6$$
where $v_{\la-n\a}$ is the unique singular vector generating the submodule
$\ver^1\cong M_{\la-n\a}.$ This implies that $\tw$ is a triangular
operator: $\tw\ver\subset\ver^1\otimes U,$ so
$$Tr|_{\ver}(\tilde\Phi_\la^u e^x)=Tr|_{\ver^1}(\tilde\Phi_\la^u
e^x).$$ Let $\tw v_{\la-n\a}=v_{\la-n\a}\otimes w +\text{\hot}.$ Using
the fact that $\ver^1\cong M_{\la-n\a}$ we see that
$$Tr|_{\ver}(\tilde\Phi_\la^u e^x)=
Tr|_{M_{\la-n\a}}(\Phi_{\la-n\a}^w e^x)=
\frac{Tr|_{M_{\la-n\a}}(\tilde\Phi_{\la-n\a}^w e^x)}
{\chi_\theta(\la-n\a)}.\tag 4-7$$

It is an easy calculation to show that
$$w=\sum_{k,l} S_{kl}^\mu(\la) g_k^\mu \omega(g_l^\mu) u=
\sum_{k,l} \( \chi_\theta(\la) (F_\mu^{-1})_{kl}\) \cdot g_k^\mu
\omega(g_l^\mu) u$$

Introduce a linear operator
$$\tilde B_n^\a(\la)=\sum_{k,l}\frac {S_{kl}^\mu(\la)} {\chi_\theta(\la-n\a)}
\cdot g_k^\mu \omega(g_l^\mu)\in End(U[0]) \tag 4-8$$

We can rewrite (4-7) as

$$\Psi(\la,x)=\Psi(\la-n\a,x) \tilde B_n^\a(\la),$$

To complete the proof we only need to show that $\tilde B_n^\a(\la)$
satisfies the required condition. It is clear that $\tilde B_n^\a(\la)$ is
rational in $\la$ and is not singular in the hyperplane $\KK=0.$
As we can rewrite

$$\tilde B_n^\a(\la)=\sum_{k,l}\frac
{\chi_\theta(\la)(F_\mu^{-1})_{kl}(\la)}{\chi_\theta(\la-n\a)}
\cdot g_k^\mu \omega(g_l^\mu)\in End(U[0]),$$

the rest follows from the Proposition 3.4.
\qed
\enddemo

Now we can introduce our main object of study. Set
$\kappa=\sum_{\a\in\DP}n_\a.$ Put
$$\psi(\la,x)=2^\kappa \Psi(\frac\la2-\rho,2x)\tag4-9$$
The properties of $\Psi$-function can now be rewritten in the
following form:

\proclaim{Corollary 4.2}
\roster
\item $\psi$-function can be represented as
$$\psi(\la,x)=e^{\<\la,x\>} P(\la,x),\tag 4-10$$
where $P(\la,x)$ is a polynomial in $\la$ of the form
$$P(\la,x)=\prod_{\a\in\DP}\<\a,\la\>^{n_\a}+\text{\ldt}.$$
\item If $\<\a,\la\>=0$ for some $\a\in\DP$, then for $s=1,2,\dots,n_\a$
$$\psi(\la+s\a,x)=\psi(\la-s\a,x)\cdot B_s^\a(\la),\tag4-11$$
where $B_s^\a(\la)=\tilde B_s^\a(\frac \la2-\rho)$ is a rational
$End(U[0])$-valued function of $\la$ and can be represented as
$$B_s^\a(\la)=b_s^\a + \text{\ldt}$$
for some constant $b_s^\a\in End(U[0]).$

\endroster
\endproclaim


\vskip .1in
\centerline{\bf 5. Uniqueness of the $\psi$-function.}
\vskip .1in

In this section we prove the uniqueness property of the function
$\psi(\la,x)$, satisfying (4-10) and (4-11).

\proclaim{Proposition 5.1}

Suppose we have an $End(U[0])$-valued function
$$\phi(\la,x)=e^{\<\la,x\>}Q(\la,x),$$
where $Q(\la,x)$ is a polynomial in $\la,$
satisfying (4-11).
Then the highest term of $Q(\la,x)$ is divisible by
$$\prod_{\a\in\DP} \<\a,\la\>^{n_\a}.$$
\endproclaim

\demo{Proof}
Consider the highest term of $Q(\la,x).$ We need
to show that it is divisible by $\<\a,\la\>^{n_\a}$ for any
$\a\in\DP.$

Fix an $\a\in\DP.$ We can uniquely represent $Q(\la,x)$ as

$$Q(\la,x)=\sum_{l=0}^L \sum_{k=0}^{K_l}\la_\a^k Q_{kl}(\lp,x),\tag5-1$$
where $Q_{kl}(\lp,x)$ are homogeneous $End(U[0])$-valued polynomials
in $\lp$ of degree $l.$

The highest term of $Q(\la,x)$ will be some combination of the
terms $\la_\a^kQ_{K_ll}(\la,x).$ We claim that it is enough to show that
$K_L\ge
n_\a.$ Indeed, it will follow then that the highest term will have
degree at least $L+n_\a,$ and therefore all terms of the form
$\la_\a^kQ_{K_ll}(\la,x)$ contributing to the highest term must have
$K_l\ge L+n_\a-l \ge n_\a,$ which proves the statement.

By our assumption $\psi(\la,x)$ satisfies (4-11), so
we can write

$$\gather
e^{s\<\a,x\>} \(\sum_{l=0}^L \sum_{k=0}^{K_l} s^k Q_{kl}(\lp,x)\)=\\
e^{-s\<\a,x\>} \(\sum_{l=0}^L \sum_{k=0}^{K_l} (-s)^k
Q_{kl}(\lp,x)\)(b_s^\a+\text{\ldt}),\tag5-2
\endgather$$
where \ldt are understood with respect to $\lp.$

We can consider homogeneous parts of (5-2) of degree $L$ in $\lp.$
Formally, given a function $f(\lp)$, we consider $$\lim_{t\to\infty}
\frac{f(t\lp)}{t^L}.$$

This gives us

$$e^{s\<\a,x\>} \(\sum_{k=0}^{K_L} s^k Q_{kL}(\lp,x)\)=
e^{-s\<\a,x\>} \(\sum_{k=0}^{K_L} (-s)^k
Q_{kL}(\lp,x)\)b_s^\a \tag5-3$$
for $s=1,\dots,K_L.$

The rest is based on the following

\proclaim{Lemma 5.2}
Consider a homogeneous system of $N$ linear equations on $K$
vector variables $A_k(z)\in\C^M$, which are meromorphic in some
additional parameter $z$:
$$\sum_{k=1}^K s^k (e^{sz}+C_s) A_k=0,\tag5-4$$
$s=1,\dots,N, C_s\in Mat_M(\C).$

If $K\le N$ then this system has only trivial solution $A_k(z)=0.$
\endproclaim
\demo{Proof of lemma}

We can think of this system as a system of linear equations on $KM$
variables $(A_k)_m$ and rewrite (5-4) in the block-matrix form

$$\matrix
{\pmatrix
(e^z\cdot Id+C_1)    & (e^z\cdot Id+C_1)    & \dots & (e^z\cdot Id+C_1)\\
2(e^{2z}\cdot Id+C_2) & 2^2(e^{2z}\cdot Id+C_2) & \dots & 2^K(e^{2z}\cdot
Id+C_2)\\
\vdots               & \vdots            &       &  \vdots          \\
N(e^{Nz}\cdot Id+C_N) & N^2(e^{Nz}\cdot Id+C_N)
& \dots & N^K(e^{Nz}\cdot Id+C_N)
\endpmatrix}
&
{\pmatrix
A_1 \\ A_2 \\ \vdots \\ A_K
\endpmatrix}
\endmatrix
=
\pmatrix 0 \\ 0 \\ \vdots \\ 0 \endpmatrix
$$

Suppose $K\le N.$ Then the determinant of the submatrix, consisting of
first $K$ blocks (or, equivalently, first $KM$ equations) is an entire
$Mat_M(\C)$-valued function of $z$ with the asymptotics as $z\to +\infty$

$$\det{\pmatrix
(e^z\cdot Id)    & (e^z\cdot Id)    & \dots & (e^z\cdot Id)\\
2(e^{2z}\cdot Id) & 2^2(e^{2z}\cdot Id) & \dots & 2^K(e^{2z}\cdot Id)\\
\vdots               & \vdots            &       &  \vdots          \\
K(e^{Kz}\cdot Id) & K^2(e^{Kz}\cdot Id) & \dots & K^K(e^{Kz}\cdot Id)
\endpmatrix},$$

which is equal to

$$\(D\cdot e^{\frac{K(K+1)}2 z} \)^M,$$

where the nonzero constant

$$D=\det
\pmatrix
1 & 1   & \dots & 1   \\
2 & 2^2 & \dots & 2^K \\
\vdots & \vdots & \ddots & \vdots \\
K & K^2 & \dots & K^K
\endpmatrix$$

is the Vandermonde determinant.

Therefore, this determinant is a nonzero entire function, which
implies that for generic $z$ it is not zero, so the
system has only trivial solution. The meromorphic functions
$A_k(z)$ are equal to zero for generic $z$, and therefore
must be identically equal to zero.

The lemma is proved.\qed
\enddemo

We now apply Lemma 5.2 to the system given by (5-3), for
$N=n_\a$, $K=K_L+1$, $M=\dim U[0]$ and setting
$C_s=(-1)^{k+1}(b_s^\a)^t, \ z=x_\a=\frac{\<\a,x\>}{\<\a,\a\>}\a$,
where $A^t$ is the transposed matrix $A$.

Consider the rows of the matrix, corresponding to $\psi(\lp,x)$, and
transpose them so that they become columns. By (5-3) all these
columns satisfy the system of equations (5-4), and as $\psi$-function
is not identically equal to zero, it implies that the system (5-4) has
a nontrivial solution. By Lemma 5.2, we have $K_L+1>n_\a$, or,
equivalently, $K_L\ge n_\a.$

The proposition is proved. \qed
\enddemo

\proclaim{Corollary 5.3}

Any $\End$-valued function
$$\phi(\la,x)= e^{\<\la,x\>} Q(\la,x),$$
satisfying (4-11), can be represented as
$$\phi(\la,x)= q(\la) \psi(\la,x)$$
for some $\End$-valued polynomial $q(\la).$
\endproclaim

\demo{Proof} The proof is similar to the proof of the Lemma in Section
1 of \cite{CV1}. We use induction on the degree of $Q(\la,x)$.

If $\deg Q(\la,x)< \deg P(\la,x)$, where
$$\psi(\la,x)=e^{\<\la,x\>}P(\la,x),$$
then by proposition we have $Q(\la,x)\equiv 0$, so
we can take $q(\la)\equiv 0.$

Suppose we have proved the statement for all polynomials
of degree less than the degree of $Q(\la,x)$.
By Proposition we can find a $\End$-valued polynomial $q_1(\la)$
such that
$$\text{highest term of }Q(\la)=
q_1(\la) \prod_\a \<\a,\la\>^{n_\a}$$
Consider the function
$$\phi_1(\la,x)=\phi(\la,x)-q_1(\la,x)\psi(\la,x).$$
Obviously, it satisfies (4-11). Moreover, it can be represented as
$$\phi_1(\la,x)=e^{\<\la,x\>}\tilde Q(\la,x),$$
where polynomial $\tilde Q(\la,x)$ has degree smaller than that of
$Q(\la,x).$

By induction hypothesis we can introduce a $\End$-valued
polynomial $q_2(\la)$ such that
$$\phi_1(\la,x)=q_2(\la)\psi(\la,x).$$
The polynomial $q(\la)=q_1(\la)+q_2(\la)$ satisfies the required
property.
\qed
\enddemo

\proclaim{Corollary 5.4}
The function $\psi(\la,x)$, satisfying both (4-10) and (4-11),
exists and is unique.
\endproclaim
\demo{Proof}
It is a direct consequence of (4-10) and Corollary 5.3.\qed
\enddemo


\vskip .1in
\centerline{\bf 6. Existence of differential operators}
\vskip .1in

The properties of the $\psi$-function obtained in Chapters 4,5 are very
close to the axioms in [CV1]. The function satisfying
these axioms was used to construct a ring of
differential operators that contained $\dim\h$ algebraically
independent operators, corresponding to the generators of the ring of
W-invariant polynomials, but was bigger than the ring generated
by those operators.

Here we apply these ideas to construct a similar ring of
matrix differential operators and thus prove
Theorem 2.6.

\proclaim{Theorem 6.1}
For any $\End$-valued polynomial $Q(\la)$ satisfying the property

$$Q(\la+s\a)=Q(\la-s\a), \qquad s=1,2,\dots,n_\a\tag6-1$$

whenever $\<\a,\la\>=0$, there exists a differential operator $D_Q$
with coefficients in $End(U[0])$, such that
$$D_Q \psi(\la,x) = Q(\la) \psi(\la,x).\tag6-2$$
The correspondence $Q(\la)\mapsto D_Q$ is a homomorphism of rings.
In particular, all $D_Q$ commute with each other.
\endproclaim

\demo{Proof}

We use induction on the degree of $Q(\la).$
If $\deg Q(\la)=0$, then $Q(\la)=const,$ so the operator
$D_Q$ will be just the operator of multiplication by this constant.

Suppose we have proved the theorem for all polynomials of
degree less than that of $Q(\la).$ Let the highest term of
$Q(\la)$ be equal to
$$\text{highest term of }(Q(\la))=\sum_\mlt a_\mlt \la^\mlt,$$
where $\mlt$ is a multiindex, $a_\mlt\in\End.$
Consider the operator $\tilde D_Q$ defined by
$$\tilde D_Q= \sum_\mlt a_\mlt \frac {\partial^{|\bold n|}
}{\partial x^\mlt}.$$
It has the property that
$$\tilde D_Q \psi(\la,x)=Q(\la)\psi(\la,x)+\text{\ldt},$$
and it also satisfies (6-1).
Consider the difference
$$\phi(\la,x)=\tilde D_Q\psi(\la,x)-Q(\la)\psi(\la,x).$$
It satisfies (4-11) and therefore can be represented as
$$\phi(\la,x)=\tilde Q(\la)\psi(\la,x)$$
for some $\End$-valued polynomial $\tilde Q(\la)$ such that
$\deg \tilde Q(\la)<\deg Q(\la).$
By induction hypothesis we can introduce an operator
$D_{\tilde Q}$ such that
$$\phi(\la,x)=D_{\tilde Q}\psi(\la,x)$$
and the operator
$$D_Q=\tilde D_Q + D_{\tilde Q}$$
has the required property, which completes the proof of the
induction step.

The assertion that the constructed correspondence is a homomorphism
of rings follows from the fact that the operator
$D_{Q_1Q_2}-D_{Q_1}D_{Q_2}$ annihilates the $\psi$-function for any $\la$, and
therefore has to be identically zero.
\qed
\enddemo

Among the polynomials $Q(\la)$, satisfying (6-1), are all $W$-invariant
polynomials $p_1(\la)$,...,$p_r(\la)$. It is known that they are
algebraically independent, and
the ring generated by corresponding differential operators is a
ring of polynomials in generators $D_{p_1},\dots,D_{p_r}.$

There are also other polynomials, satisfying (6-1), which are not
$W$-invariant. They give rise to differential operators which are
not $W$-invariant and therefore do not belong to the ring generated
by $D_{p_1},\dots,D_{p_r}.$

In particular, all polynomials contained in the ideal generated by
$$Q_0(\la)=\prod_{\a\in\DP} \prod_{n=1}^{n_\a}\(\<\a,\la\>^2-n^2\<\a,\a\>^2\)$$
satisfy the (6-1).

\proclaim{Proposition 6.2} The differential operator corresponding to
the invariant polynomial $p_1(\la)=\<\la,\la\>$, is equal to
$$D_{p_1}=\Delta_\h - \sum_{\a\in\DP} \frac{2e_\a
f_\a}{sinh^2\<\a,x\>}\tag 6-3.$$
\endproclaim

This fact can be proved by a direct computation.
Another proof of it using the relationship
between the center of $U(\g)$ and commuting differential operators
is sketched in the Appendix (see also \cite{E}).

The operator (6-3) coincides with the
 generalized Calogero-Sutherland operator (2-3) for the trigonometric
case. We have shown that this operator is algebraically integrable.
Theorem 2.6 is proved.

\vskip .1in

\centerline{\bf Appendix}
\vskip .1in

In conclusion we briefly describe how to construct
quantum integrals of $H_{\g,U,u}$
with trigonometric potential from central elements
of $U(\g)$. This construction works for arbitrary
module $U$, not necessarily highest weight.

\proclaim{Proposition A1}
\cite{E} Let $X\in U(\g)$ be an element of degree $0$, i.e.
$[h,X]=0$, $h\in\frak h$.
Then there exists a unique matrix differential
operator $\Cal D(X)$ with $\End$-valued coefficients such that
for any $\la\in \frak h^*$ and any intertwining operator
$\Phi: M_\la\to M_\la\otimes U$
$$
{Tr|_{\ver}(\Phi X
e^x)} =\Cal D(X)Tr|_{\ver}(\Phi
e^x).
$$
\endproclaim

The proof of this theorem and a recursive construction of $\Cal D(X)$
is given in \cite{E}. We illustrate the idea of this construction
by computing $D(EF)$ for $\g=\frak{sl}_2$ ($E,F,H$ are standard
generators of $\g$). The main trick is to carry $E$ around the trace,
using the intertwining property of $\Phi$ and the
cyclic property of the trace:
$$
\gather
Tr|_{\ver}(\Phi EF
e^x)=Tr|_{\ver}((E\otimes 1)\Phi F
e^x)+E Tr|_{\ver}(\Phi F
e^x)=\\ Tr|_{\ver}(\Phi F
e^xE)+E Tr|_{\ver}(\Phi F
e^x)=
e^{<\alpha,x>}Tr|_{\ver}(\Phi FE
e^x)+E Tr|_{\ver}(\Phi F
e^x)=\\ e^{<\alpha,x>}Tr|_{\ver}(\Phi (EF+H)
e^x)+E Tr|_{\ver}(\Phi F
e^x)=\\
e^{<\alpha,x>}Tr|_{\ver}(\Phi EF
e^x)+e^{<\alpha,x>}\frac{\dd}{\dd\alpha}Tr|_{\ver}(\Phi
e^x)+E Tr|_{\ver}(\Phi F
e^x),\tag A1\endgather
$$
(where $\alpha$ is the positive root of $\g$) which implies
$$
Tr|_{\ver}(\Phi EF
e^x)=\frac{1}{1-e^{<\alpha,x>}}\biggl(e^{<\alpha,x>}\frac{\dd}
{\dd\alpha}Tr|_{\ver}(\Phi
e^x)+E Tr|_{\ver}(\Phi F
e^x)\biggr).\tag A2
$$
Further, we have
$$
Tr|_{\ver}(\Phi F
e^x)=Tr|_{\ver}((F\otimes 1)\Phi
e^x)+FTr|_{\ver}(\Phi
e^x)=e^{-<\alpha,x>}Tr|_{\ver}(\Phi
Fe^x)+FTr|_{\ver}(\Phi
e^x),\tag A3
$$
which implies
$$
Tr|_{\ver}(\Phi Fe^x)=\frac{1}{1-e^{-<\alpha,x>}}
FTr|_{\ver}(\Phi e^x).\tag A4
$$
Combining (A2) and (A4), we find
$$
\gather
Tr|_{\ver}(\Phi EF
e^x)=\\
\biggl(\frac{e^{<\alpha,x>}}{1-e^{<\alpha,x>}}\frac{\dd}{\dd\alpha}
+\frac{1}{(1-e^{<\alpha,x>})(1-e^{-<\alpha,x>})}
EF\biggr)Tr|_{\ver}(\Phi e^x).\tag A5 \endgather
$$
Thus
$$
\Cal D(EF)=
\frac{e^{<\alpha,x>}}{1-e^{<\alpha,x>}}\frac{\dd}{\dd\alpha}
+\frac{1}{(1-e^{<\alpha,x>})(1-e^{-<\alpha,x>})}
EF.\tag A6
$$

\vskip .1in

In general, it is not true that $\Cal D(X_1X_2)$ equals either
$\Cal D(X_1)\Cal D(X_2)$ or $\Cal D(X_2)\Cal D(X_1)$. However:

\proclaim{Proposition A2}\cite{E} If $X_1$ belongs to the center of
$U(\g)$, then for any $X_2$ one has $\Cal D(X_2)\Cal D(X_1)=\Cal D(X_1X_2)$.
\endproclaim

This proposition follows from the fact that if $\Phi$ is an
intertwining operator then $\Phi X_1$ is also an
intertwining operator.

\proclaim{Corollary} If $X_1,X_2$ are both in the center of $U(\g)$, then
$\Cal D(X_1)$ and $\Cal D(X_2)$ commute with each other.
\endproclaim

Let $D(X)$ be the differential operator obtained from
$\Cal D(X)$ by conjugation by the function $Tr|_{M_{-\rho}}(e^x)$,
i.e. defined by
$D(X)\xi(x)=Tr|_{M_{-\rho}}(e^x)\Cal
D(X)(\frac{\xi(x)}{Tr|_{M_{-\rho}}(e^x)})$.
The following statement is checked by a direct computation:

\proclaim{Proposition A3}\cite{E} Let $B$ be an orthonormal basis of
$\g$, and $C_1=\sum_{a\in B}a^2$ be the Casimir element. Then
$$
D(C_1)=\Delta_{\frak
h}-\sum_{\alpha\in\Delta^+}\frac{e_{\alpha}f_{\alpha}}{2sinh^2(<\alpha,x>/2)}
+\text{const}.\tag A7
$$
\endproclaim

\proclaim{Proposition A4}
Let $C_1,...,C_m$ be algebraically independent generators
of the center of $U(\g)$. Let $\hat L_i=D(C_i)$. Then
$\hat L_1,...,\hat L_m$ are algebraically independent
quantum integrals of the Schr|''odinger operator
$\hat L_1$ given by (A7). The symbols of $\hat L_i$ generate
the algebra of Weyl group invariant polynomials on $\frak h$.
The function $\Psi(\la,x)$ is a joint eigenfunction
for $\hat L_1,...,\hat L_m$ with eigenvalues $\phi(C_i)(\la)$,
where $\phi$ is the Harish-Chandra homomorphism defined in Section 3.
\endproclaim

Observe that
 operator (A7) transforms into $\frac{1}{4}H_{\g,U,u}+\text{const}$ when
one makes a change of variables $x\to 2x$. Therefore, we have

\proclaim{Proposition A5} Let $L_i$ be the operators
obtained from $\hat L_i$ by replacing $x$ with $2x$.
Then $L_1=\frac{1}{4}H_{\g,U,u}+\text{const}$, and $L_1,...,L_m$ are
algebraically independent quantum integrals of $H_{\g,U,u}$ for
trigonometric $u$. The function $\psi(\la,x)$ is a joint eigenfunction
of $L_1,...,L_m$.
\endproclaim

This implies Theorem 2.3 in the trigonometric case.

Finally, we observe that if $C$ is a central element of $U(\g)$ then
by Propositions A4,A5 one has $D(C)=D_p$, where $D_p$ is defined in Section 6,
and $p$ is the Weyl group invariant polynomial given by $p(\la)=
\phi(C)(2(\la+\rho)$. This proves Proposition 6.2.

\end